\documentclass{ar-1col-S2O}
\usepackage[numbers]{natbib}
\usepackage{url}
\usepackage{comment}
\usepackage{graphicx}
\usepackage{amsmath,bm}
\usepackage{amssymb}
\usepackage{setspace}
\usepackage[ngerman,greek,english]{babel}
\usepackage{upgreek}
\usepackage{todonotes}
\usepackage{pdfpages}
\usepackage{footmisc}
\usepackage{enumitem}
\usepackage{mathtools}
\usepackage{hyperref}
\usepackage{tabularx}
\usepackage{longtable}
\usepackage{siunitx}
\usepackage{braket}

\newcommand{\be}{\begin{equation}}
\newcommand{\ee}{\end{equation}}

\usepackage{ulem}
\newcommand\redsout{\bgroup\markoverwith{\textcolor{red}{\rule[0.5ex]{2pt}{1.4pt}}}\ULon}

\begin{document}

\markboth{Bonaiti et al.}{Constraining Nuclear Matter from Finite Nuclei}

\title{Constraining Nuclear Matter from Finite Nuclei: the Ab Initio Perspective}

\author{F. Bonaiti,$^{1,2}$, 
S. Bacca,$^{3,4}$, and I. Tews $^5$
\affil{$^1$Facility for Rare Isotope Beams, Michigan State University, East Lansing, USA, MI 48824; email: bonaiti@frib.msu.edu}
\affil{$^2$Physics Division, Oak Ridge National Laboratory, Oak Ridge, USA, TN 37831}
\affil{$^3$ Institut f\"ur Kernphysik and PRISMA$^+$ Cluster of Excellence, Johannes Gutenberg-Universit\"at, Mainz, Germany, D-55128}
\affil{$^4$Helmholtz-Institut Mainz, Johannes Gutenberg-Universit\"at Mainz, Mainz, Germany, D-55099;  email: s.bacca@uni-mainz.de}
\affil{$^5$Theoretical Division, Los Alamos National Laboratory, Los Alamos, New Mexico 87545, USA;  email: itews@lanl.gov}
}


\begin{abstract}
This review surveys recent developments in ab initio calculations of finite nuclei aimed at constraining the nuclear matter equation of state. 
Particular emphasis is placed on observables that connect finite-nucleus structure to bulk properties of nuclear matter, including neutron-skin thicknesses, electric dipole polarizabilities, and monopole sum rules.
We discuss how neutron skins and dipole polarizabilities provide information on the density dependence of the symmetry energy, while monopole observables constrain the incompressibility of nuclear matter. 

\end{abstract}

\maketitle

\tableofcontents

\section{INTRODUCTION}

Nuclear matter is the densest form of matter that we can directly observe in the universe. 
It is described as an infinite system of nucleons, neutrons and protons, that interact via strong nuclear forces. 
When the number of neutrons and protons is the same, the system is called symmetric nuclear matter and can be found in the center of heavy atomic nuclei.
In these systems, nuclear matter can reach densities of $n_0 \approx 0.16$ fm$^{-3}$, the so-called nuclear saturation density, which corresponds to a mass density of $\sim 2.7\cdot 10^{14}$g/cm$^3$.
This is roughly 100 trillion times the density of water.
When nuclear matter consists mostly of neutrons, it is called neutron-rich matter and can be found in the cores of neutron stars, reaching even higher densities of several times $n_0$. 
The properties of these systems are determined by the properties of the nuclear interactions between its constituents.

Neutron stars offer the opportunity to study dense nuclear matter at densities exceeding what can be realized in experiments on Earth. 
These fascinating astrophysical objects originate from the supernova explosions of stars with total masses between around $8$ and $20$ solar masses~\cite{lattimer2004}. 
The remnants of these explosions, neutron stars, are typically characterized by a mass of 1.4-2.0 times the mass of our sun, $M_{\odot}$, squeezed into a sphere with a radius between 10 and 14 km. 
The physics of neutron stars is determined by the competition between gravity and the quantum mechanical pressure produced by its interacting constituents (mostly neutrons, with small fractions of protons and electrons), precluding the collapse of the star into a black hole.
The study of neutron stars and its neutron-rich matter is, hence, inherently connected to exploring the strong nuclear interaction between nucleons, which acts over distances of the order of $1$~fm. 
Consequently, investigating the structure of neutron stars creates a bridge between areas of physics that differ by over fifteen orders of magnitude in length scales but are connected by quantum mechanics.

A variety of theoretical, experimental, and observational approaches using data originating both in terrestrial laboratories and in space, can be employed to constrain the equation of state (EOS) of dense nuclear matter across various density regimes (see Ref.~\cite{piekarewicz2022_ns,Koehn:2024set} and references therein). 
A remarkable example of the synergy between these different techniques was the groundbreaking detection of the gravitational-wave event GW170817~\cite{ligo}, which inaugurated the ``golden era" of multimessenger astronomy~\cite{ligo3}. 
Attributed to a binary neutron-star merger, this event enabled the determination of nuclear-matter properties and its EOS by analyzing the tidal deformabilities of the neutron stars during the inspiral phase of their merger~\cite{ligo2}.
In this review, we concentrate on constraints for the EOS near the characteristic density found within atomic nuclei — the nuclear saturation density $n_0\approx 0.16$ fm$^{-3}$ —  at which the nuclear matter EOS represents a reasonable approximation for the conditions found in the outer core of neutron stars~\cite{piekarewicz2022_ns}.

In addition to neutron stars, properties of finite nuclei, can also be used to constrain the EOS around saturation density. 
A prominent example is the neutron-skin thickness defined as the difference between the neutron and proton radii in a nucleus.
Theoretical studies have revealed correlations between neutron skins in neutron-rich nuclei and key properties of the EOS~\cite{centelles2010,rocamaza2015}, which can be leveraged in the presence of precise experimental measurements of this observable. 
The latter, however, present significant challenges (see Ref.~\cite{thiel2019} for a recent review). 
As the neutron has no electric charge, the neutron distribution has been tested in the past via hadronic probes, where the treatment of strong interactions among nucleons in the analysis is a potential source of systematic uncertainties.
Limiting such uncertainties is work in progress~\cite{tsaran2025}.  
At the moment, parity violating elastic electron scattering (PVES), relying on the electroweak interaction, has emerged as as a cleaner technique to extract the neutron-skin thickness~\cite{donnelly1989}. 
In the last few years, it has been used to measure neutron skins in $^{48}$Ca~\cite{crex} and $^{208}$Pb~\cite{prex1,prex}.
The analysis and implications of these measurements on the EOS are still under discussion in the nuclear physics and astrophysics communities~\cite{reed2021,essick2021,reinhard2022,hu2022,reed2024}.

Another strategy to constrain the EOS using experiments with finite nuclei uses electromagnetic observables, which describe how nuclei respond to interactions with photons. 
The density dependence of the EOS can be explored by slightly exciting the nucleus from its ground state using electromagnetic probes. 
In this case, correlations have been identified between key EOS parameters and the electric dipole polarizability, $\alpha_D$~\cite{rocamaza2013}, which quantifies how easily the nuclear charge distribution is distorted by an external electromagnetic field. 
In addition, other EOS properties, such as the incompressibility of nuclear matter, can be constrained through systematic studies of collective vibrations like isoscalar monopole resonances~\cite{blaizot1980}.
From a theoretical standpoint, extensive investigations of electromagnetic observables across the nuclear chart have primarily been performed within the energy density functional (EDF) framework, often combined with the random phase approximation (RPA) and its extensions (see, e.g., Refs.~\cite{paar2007,rocamaza2018,colo2022} and references therein). 
In this review, however, we focus on describing these observables within an \textit{ab initio} framework.

This review is organized as follows. In Sections~\ref{eos} and \ref{abinitio}, we introduce the nuclear matter EOS and ab initio nuclear theory, respectively. 
In Section~\ref{symmetry_energy}, we discuss constraints on the symmetry energy, with separate subsections devoted to the neutron-skin thickness and the electric dipole polarizability. 
In Section~\ref{k-constraints-section}, we present constraints on the nuclear incompressibility. 
Finally, in Section~\ref{conclusions}, we summarize our conclusions.

\section{THE NUCLEAR MATTER EQUATION OF STATE}
\label{eos}
\subsection{Dense-Matter Modeling}
\label{nuclear-eos-section}

We now describe how to model the EOS of dense nuclear matter at densities up to about $1-2$ times the saturation density, where meaningful constraints from nuclear physics can be established.
We will also focus on the EOS at zero temperature, as the connection to ground-state properties of nuclei is temperature independent.
Furthermore, although the temperature in neutron stars can reach $T\sim10^8$~K, matter can be treated in the zero-temperature approximation because the corresponding thermal energy, $k_B T\sim 10$~keV, is much smaller than the characteristic nuclear energy scales of the order of 10~MeV.
Under these assumptions, we can describe nuclear matter as an idealized, cold, uniform system of neutrons and protons interacting solely via strong nuclear forces. 

\begin{figure}[t]
\centering
    \includegraphics[scale=0.15]{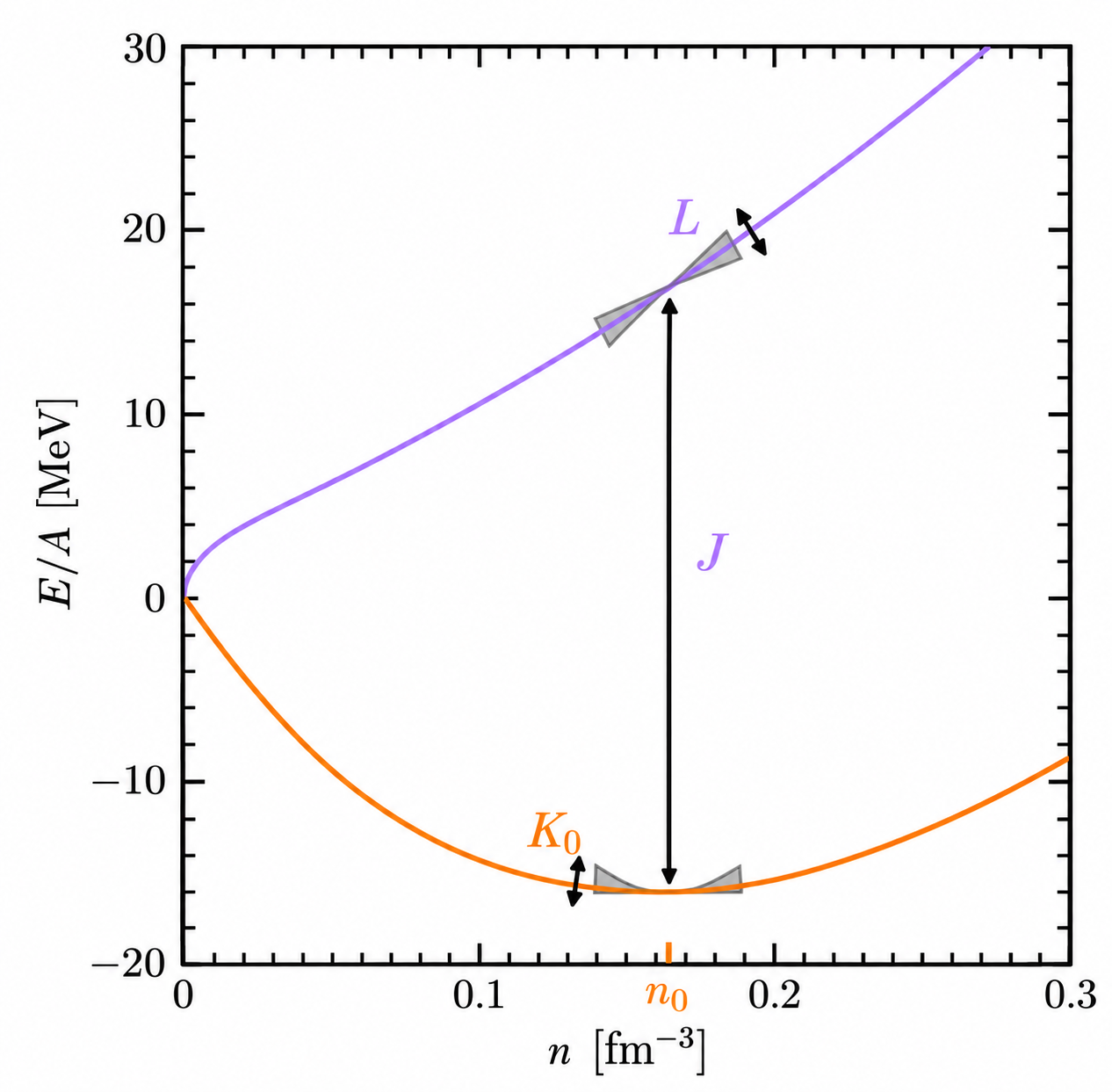}    
    \caption[Equation of state of pure neutron matter and symmetric nuclear matter]{Equation of state of pure neutron matter (purple) and symmetric nuclear matter (orange) as functions of baryon number density $n$. 
    The parameters defining the EOS at saturation density $n_0$, corresponding to the energy per nucleon of SNM at saturation, incompressibility $K_0$, symmetry energy at saturation $J$, and slope parameter $L$ are also highlighted.}
    \label{fig:snm-pnm-eos}
\end{figure}

We describe the system by an infinite number of nucleons, $A$, in an infinite volume, $V$, such that the total number density $n = A/V$ remains finite. 
The asymmetry parameter  
\begin{equation}
\alpha = \frac{n_n - n_p}{n}\,,
\end{equation}
quantifies the difference between the neutron density $n_n$ and the proton densities $n_p$. 
Consequently, $n = n_p + n_n$ is the total nucleon number density. 
For $\alpha = 1$, the system corresponds to pure neutron matter (PNM).
In this case, both the energy per particle and the pressure are positive, indicating that neutrons alone do not form self-bound states.
In contrast, symmetric nuclear matter (SNM) is characterized by equal neutron and proton densities ($\alpha = 0$). 
In SNM, the energy per particle exhibits a minimum at the saturation density $n_0$, indicating the existence of a stable bound state and leading to zero pressure at that specific density.
The energy per particle $E/A$ as a function of density for SNM and PNM is shown in Figure~\ref{fig:snm-pnm-eos}.

The nuclear symmetry energy is usually defined as the difference between the energy per nucleon of PNM and SNM 
\begin{equation}
    S(n) = \frac{E}{A}(n, 1) - \frac{E}{A}(n, 0)\,,
\end{equation}
where $\frac{E}{A}(n, 0)$ is the EOS of SNM and $\frac{E}{A}(n, 1)$ is the EOS of PNM.
Therefore, the symmetry energy measures the energy cost required to replace protons with neutrons in dense matter.
Generally, the dependence of $E(n, \alpha)$ for arbitrary $\alpha$ is modeled as
\begin{equation}
    \frac{E}{A}(n, \alpha) = \frac{E}{A}(n, 0) + S_2(n) \alpha^2 + \mathcal{O}[\alpha^4]\,.
\label{infmateos}
\end{equation}
If this quadratic approximation holds, then $S_2(n)$ can be identified with the symmetry energy.
As shown by various nuclear many-body calculations (see for example Ref.~\cite{Somasundaram:2020chb} and references therein), truncating Eq.~(\ref{infmateos}) at quadratic contributions in $\alpha$ provides a good approximation of the dependence of $E/A$ on $\alpha \in [0,1]$ for densities up to $2 n_0$.  
Furthermore, odd powers of $\alpha$ are absent in a polynomial expansion due to the isospin invariance of nuclear interactions. 

By expanding the EOS of SNM around $n_0$, one obtains
\begin{equation}
    \frac{E}{A}(n, 0) = \frac{E}{A}(n_0, 0) + \frac{1}{2} K_0 \left(\frac{n - n_0}{3n_0}\right)^2 + \mathcal{O}(n - n_0)^3\,,
    \label{eq:SNM}
\end{equation}
where $\frac{E}{A}(n_0, 0)$ is the energy per nucleon of SNM at saturation density. 
The pressure $P = n^2 \frac{\partial E/A}{\partial n}$ vanishes at nuclear saturation density, which explains why there is no term linear in $n$ in Eq.~(\ref{eq:SNM}).
We define the incompressibility of SNM as 
\begin{equation}
    K_0 = 9 \frac{\partial P}{\partial n} \bigg|_{n_0, \alpha = 0} = 9 n_0^2 \frac{\partial^2 E/A}{\partial n^2} \bigg|_{n_0, \alpha = 0}\,.
    \label{incompressibilitysnm}
\end{equation}
As shown in Figure~\ref{fig:snm-pnm-eos}, $K_0$ describes the curvature of the energy per particle of SNM around saturation density. 
It measures the resistance of  nuclear matter to compression and expansion. 

The symmetry energy $S(n)$ can also be expanded around $n_0$. 
In this case we get 
\begin{equation}
        S(n) = J + L \frac{n - n_0}{3n_0} + \frac{1}{2} K_{\rm sym} \left(\frac{n - n_0}{3n_0}\right)^2 + \mathcal{O}(n - n_0)^3\,,
    \label{symenergy}
\end{equation}
where $J = S(n_0)$ is the symmetry energy at saturation density, $L$ is the so-called slope parameter, and $K_{\rm sym}$ is the incompressibility (or curvature) of the symmetry energy. 
Since in SNM the pressure vanishes at saturation, the slope parameter $L$ is directly related to the pressure of PNM at saturation density,
\begin{equation}
    L = 3n_0 \frac{\partial S}{\partial n} \bigg|_{n_0} = \frac{3}{n_0} P(n_0, \alpha = 1)\,.
\end{equation}
Higher values of $L$ imply a steeper PNM curve in Figure~\ref{fig:snm-pnm-eos}. 
The slope parameter, describing the pressure among neutrons, plays a key role in interpreting several observables in neutron-rich systems, such as the neutron-skin thickness in lead or the radius of neutron stars.
For example, larger $L$ values imply a stiffer EOS, i.e., an EOS with larger pressure at a given density, which translates to larger radii of neutron stars.  
In the following, we will focus on the symmetry energy parameters $J$ and $L$ and the incompressibility $K_0$, which determine the density dependence of the EOS at and around saturation density. 

\subsection{Dense Matter in Nature: Neutron Stars}

Neutron stars explore nuclear systems over a wide  range of densities and pressures in their interiors, from atomic nuclei arranged in a lattice in the neutron-star crust at low densities to neutron-rich matter in their cores.  
Our understanding of neutron stars depends on our ability to describe these different systems.

The existence of neutron stars was first hypothesized in the 1930s. 
At that time, Landau proposed the idea of compact stars similar to giant atomic nuclei~\cite{yakovlev2013}. 
The 1933 proceedings of the American Physical Society contain the first written trace of such conjecture. 
There, Baade and Zwicky claimed: \textit{With all reserve we advance the view that supernovae represent the transition from ordinary stars into neutron stars, which in their final stages consist of extremely closed packed neutrons}~\cite{baade1934}. 
Later, in 1939, the equations governing the structure of neutron stars, the so-called Tolman-Oppenheimer-Volkoff (TOV) equations, were derived independently by Tolman~\cite{tolman1939} and Oppenheimer and Volkoff~\cite{oppenheimer1939}, employing Einstein's theory of general relativity. 
The EOS of neutron star matter is a key ingredient of the TOV equations, as it defines the pressure that counteracts gravity across the vast range of densities present within the star. 
In their landmark study~\cite{oppenheimer1939}, Oppenheimer and Volkoff, assuming that neutron-star matter behaves as a gas of non-interacting neutrons, concluded that a neutron star would collapse into a black hole if its mass exceeded approximately 0.7 $M_{\odot}$.
However, this prediction was challenged by astronomical observations, starting with Jocelyn Bell's pioneering work~\cite{hewish1968}, with recent data indicating that this threshold mass limit might be 3 times higher~\cite{Antoniadis:2013pzd,NANOGrav:2017wvv,Fonseca:2021wxt}. 
This highlights the essential role played by nuclear interactions in determining the EOS of neutron stars. 
In fact, neutron stars are stabilized by the short-range repulsive interaction between nucleons and potentially by the emergence of exotic phases of matter in their cores.

The surface of a neutron star is covered by a thin atmosphere, primarily composed of hydrogen and helium.  
Although this layer is relatively thin, it crucially affects  astrophysical observations. 
For example, the thermal radiation emitted from the atmosphere provides information on the chemical composition and other properties of the star~\cite{piekarewicz2022_ns}.  
Beneath the atmosphere lies the outer crust, where densities range from $10^4$ to $10^{11}$ \SI{}{g/cm^3}~\cite{baym1971}.  
The crust is formed by a lattice of increasingly neutron-rich nuclei embedded in a uniform electron gas. 
At densities of around $4 \times 10^{11}$ \SI{}{g/cm^3}, nuclei in the crust can no longer bind additional neutrons, and begin to be surrounded by a neutron gas that can exhibit fascinating effects due to superfluidity. 
This neutron drip density marks the boundary between the outer and inner crust.
In the inner crust, the interplay between the short-range strong interaction and the long-range Coulomb repulsion, referred to as ``Coulomb frustration", is believed to give rise to complex structures with diverse topologies, collectively known as ``nuclear pasta" phases~\cite{ravenhall1983}.  
The crust–core transition occurs at approximately half of nuclear saturation density, where nuclei dissolve completely and matter consists primarily of a uniform fluid of neutrons, with smaller fractions of protons, electrons, and muons in chemical equilibrium.  
Toward the stellar center, at densities well above $2 \rho_0$, more exotic forms of matter, such as hyperons or deconfined quarks, may appear~\cite{page2006}.
For a more detailed discussion of neutron star structure, the reader is referred to, e.g., Refs.~\cite{lattimer2004,piekarewicz2022_ns}.  

The EOS describes the thermodynamic properties of the neutron star throughout all of these layers. 
To extract structural properties of the stars, one solves the TOV equations,
\begin{equation}
\begin{split}
    &\frac{dP(r)}{dr} = -\frac{Gm(r)}{r^2}  \left( 1 + \frac{P(r)}{\rho(r)c^2}\right) \left(1 + \frac{4\pi r^3 P(r)}{m(r) c^2}\right) \left(1 - \frac{2Gm(r)}{rc^2}\right)^{-1}, \\
     &\frac{dm(r)}{dr} = 4\pi r^2 \rho(r)\,,
    \label{tov}
\end{split}
\end{equation}
which determine the behavior of a spherically symmetric body of isotropic material in hydrostatic equilibrium in general relativity. 
In Eq.~(\ref{tov}), $P(r)$ corresponds to the pressure as a function of the radial coordinate $r$, $m(r)$ defines the mass of the object contained within a given radius $r$, $c$ is the speed of light, and $\rho(r) = \epsilon(r)/c^2$ is given in terms of the energy density $\epsilon(r)$. 
The EOS, $P = P(\epsilon)$, links the pressure to the energy density and closes this set of differential equations.  
The system is then solved by imposing the boundary conditions $P = P_c$ and $m = 0$ at the center of the star ($r = 0$), and integrating outward until the pressure drops to zero, $P = 0$, which defines the stellar radius $R$ and total mass $M$ from the conditions $P(R) = 0$ and $M = m(R)$.  
For a given EOS, the total mass $M(P_c)$ and radius $R(P_c)$ can be calculated for each central pressure $P_c$, producing a unique mass--radius relation characteristic of neutron stars.  
Thus, the TOV equations serve as an efficient mapping between the equation of state $P(\epsilon)$ and the corresponding mass--radius relation $M(R)$. 
In this sense, they fully specify the structure of neutron stars.
EOS with a lower pressure at a given energy density are classified as soft, and result in smaller neutron-star radii and lower maximum masses compared to stiffer EOS that have higher pressure at a given density.   
We want to stress that softness and stiffness are density-dependent concepts, i.e., an EOS can be soft in some density range and stiff in another. 
Combined mass-radius measurements from the Neutron Star Interior Composition Explorer (NICER) (see, e.g., Refs.~\cite{Miller:2026aju}) and observations of gravitational waves from the events GW170817~\cite{ligo} and GW190425~\cite{LIGOScientific:2020aai} associated with binary neutron star mergers, provide valuable constraints.  
Since the TOV equations establish a one-to-one correspondence between the EOS and the mass-radius relation~\cite{lindblom1992}, they can be used either to predict $M(R)$ from our knowledge of the nuclear interaction or to extract information about the EOS from astrophysical data.
Constraining the EOS from both nuclear theory and neutron star observations is an active area of research (see, e.g., Refs.~\cite{Annala:2017llu,miller2020,raaijmakers2020,Dietrich:2020efo,Landry:2020vaw,huth2022,tsang2024,Somasundaram:2024ykk} for some recent works on the topic).

\section{AB INITIO NUCLEAR THEORY}\label{abinitio} 

Ref.~\cite{ekstrom2020} defines ab initio methods as \textit{a systematically improvable approach for quantitatively describing nuclei using the finest resolution scale possible while maximizing its predictive capabilities}. 
In principle, the finest resolution scale would imply a description in terms of fundamental degrees of freedom, namely quarks and gluons, whose strong interactions are described by Quantum Chromodynamics (QCD). 
At high energies, QCD is asymptotically free, making the theory perturbative. 
However, when going to low energies relevant in nuclear physics, the strong coupling constant rapidly increases, confining quarks into colorless objects such as mesons and baryons. 
The large value of the strong coupling constant makes perturbation theory not applicable in this regime, requiring non-perturbative methods and hindering a direct calculation of the properties of nuclear systems directly from QCD. 
Direct solutions within the framework of lattice QCD are possible but their predictive power is currently restricted to single mesons, baryons, and few-nucleon systems, due to their huge computational cost~\cite{Drischler:2019xuo}.

As a consequence, it is common to describe nuclear systems by choosing effective degrees of freedom, such as nucleons and pions, whose effective interactions can be described by models of nuclear forces. 
A modern framework to model nuclear forces is given by chiral effective field theory (EFT) (see, e.g.,  Refs.~\cite{epelbaum2009,machleidt2011,hammer2020} and references therein).
EFTs exploit the presence of a separation of scales, represented by an energy gap between a soft scale $Q$ and a hard scale $\Lambda_b$, with $Q \ll \Lambda_b$. 
The soft scale is the typical energy or momentum scale of the physical systems described by the EFT, while the hard scale, also called the breakdown scale, identifies the regime not explicitly resolved by the theory. 
In the case of low-energy nuclear physics, the hadron spectrum clearly shows a separation of scales between the mass of the pions, $M_{\pi} = 140$ MeV, and the mass of the next heavier mesons, $\rho$ and $\omega$, with $M_{\rho} = 770$ MeV and $M_{\omega} = 782$ MeV, respectively.
Additionally, typical nucleon momenta in nuclear systems are of the order of the pion mass.
We can then naturally identify the soft scale with $Q \sim M_{\pi}$ and the hard scale with $\Lambda \sim M_{\rho}$. 

Once a separation of scales has been identified, one writes down the most general Lagrangian consistent with the symmetries of the underlying microscopic theory, given by QCD in this case. 
Among the relevant QCD symmetries~\cite{Piarulli:2019cqu}, the explicitly and spontaneously broken chiral symmetry leads to the appearance of pions as pseudo Goldstone bosons. 
Chiral EFT potentials are then derived by systematically expanding this general Lagrangian in powers of $Q/\Lambda_b$, where contributions proportional to $(Q/\Lambda_b)^{\nu}$ enter at progressively higher orders $\nu$. 
Choosing how to connect the diagrammatic representations of the terms in the expansion to $\nu$ determines the importance of each term and defines the so-called power counting scheme of the theory. 
While different choices are under discussion by the nuclear-physics community (see e.g.~\cite{kaplan1998,yang2021,thim2023}), the most widely used power-counting scheme is the one defined by Weinberg~\cite{weinberg1990}.

Nuclear forces from chiral EFT generally contain two types of interaction vertices: contact interactions between nucleons and pion-nucleon interactions. 
Contact interactions capture short-range effects of nuclear forces, while pion-nucleon interactions lead to one- and multiple-pion exchange processes, defining the long- and medium-range behavior of nuclear forces. 
For each vertex in chiral EFT, the high-energy information integrated out from the underlying theory appears in the form of low-energy constants (LECs), that set the strength of the vertex and are commonly fitted to experiment. 
A key feature of chiral EFT is the natural appearance of many-body forces in the theory. 
At $\nu=3$ or next-to-next-to-leading order (N$^2$LO), one finds the first contributions to  three-nucleon forces, accounting for short- and medium-range many-body effects.
The inherent systematic improvability of chiral EFT, given by its order-by-order expansion, offers a significant benefit: it enables the calculation of nuclear observables with a desired level of precision, while also allowing for the estimation of uncertainties stemming from the neglected higher orders in the chiral expansion~\cite{Epelbaum:2014efa,drischler2020,Drischler:2020yad,Armstrong:2025tza,Curry:2025pna}. 

In the past 20 years, a broad range of nuclear interactions from chiral EFT have been developed, differing in the inclusion of the $\Delta$ isobar as a degree of freedom, the order at which they have been derived, the LEC fitting procedure, and the choice of regulator scheme.
The availability of these systematically improvable chiral interactions has enabled significant progress in \textit{ab initio} nuclear-structure calculations~\cite{hergert2020,ekstrom2020}. 
Although for many years \textit{ab initio} methods were limited to light nuclei, they have recently seen an exponential growth in their applicability to heavy systems up to $^{208}$Pb and $^{266}$Pb~\cite{hagen2016b,morris2018,stroberg2021,hu2022,door2025,bonaiti2025}. 
Such progress has also been fueled by advances in high-performance computing and by the development of many-body methods employing controlled approximations and scaling polynomially with mass number. 
Example of \textit{ab initio} methods are Coupled-Cluster (CC) theory~\cite{hagen2014}, In-Medium Similarity Renormalization Group (IMSRG)~\cite{hergert2016}, Lattice Effective Field Theory~\cite{lee2009}, No-Core Shell Model~\cite{barrett2013}, Quantum Monte Carlo~\cite{carlson2015}, Self Consistent Green's function (SCGF)~\cite{barbieri2018}, and many others (see Ref.~\cite{hergert2020} for a recent review on ab initio techniques). 
Similarly, these methods have been very successful in describing the properties of dense nuclear matter.

\section{CONSTRAINTS ON THE SYMMETRY ENERGY} \label{symmetry_energy}

Chiral EFT calculations of dense nucleonic matter provide theoretical constraints on the symmetry energy at saturation density $J$ and its slope parameter $L$ introduced before.
$J$ and $L$ can also be constrained through studies of ground- and excited-state properties of finite nuclei and from astrophysical observations of neutron stars. 
Hence, these parameters provide a direct link between nuclear-physics theory, experiment, and astrophysics.

In this section, we give an overview of recent theoretical and experimental efforts in this direction, focusing on the constraints provided by ab initio calculations. 
Two key observables will be discussed: the neutron-skin thickness and the electric dipole polarizability.
 
\subsection{Neutron-skin thickness}
The neutron-skin thickness of an atomic nucleus is  the difference between its point-neutron and point-proton radii,
\begin{equation}
    R_{\rm skin} = R_n - R_p\,,
    \label{nskin}
\end{equation}
and consequently, it depends on the proton and neutron density distributions.

If neutron number $N$ and proton number $Z$ are the same, $N=Z$, the proton and neutron distributions are similar, with slight differences induced by Coulomb interactions.
However, when the neutron-to-proton ratio increases, the neutron density distribution spreads out beyond  that of the protons, resulting in the formation of a neutron skin.

\begin{figure}[t]
\centering
\begin{minipage}{0.48\textwidth}
\centering
\includegraphics[width=\textwidth]{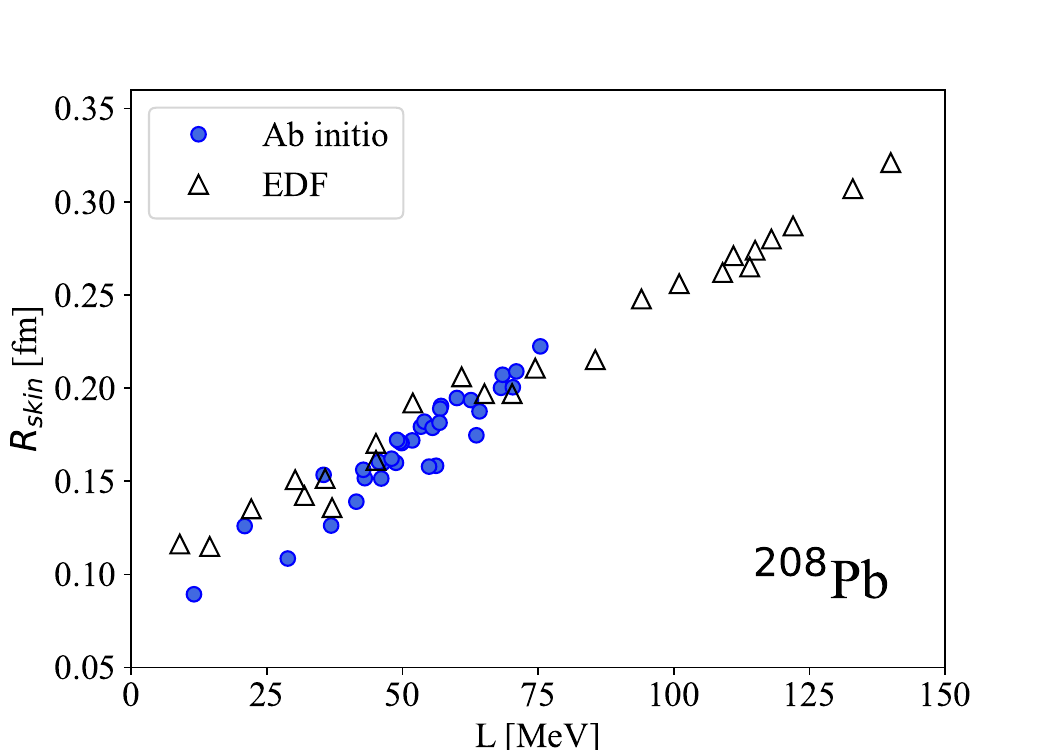}
\end{minipage}
\hfill
\begin{minipage}{0.485\textwidth}
\centering
\includegraphics[width=\textwidth]{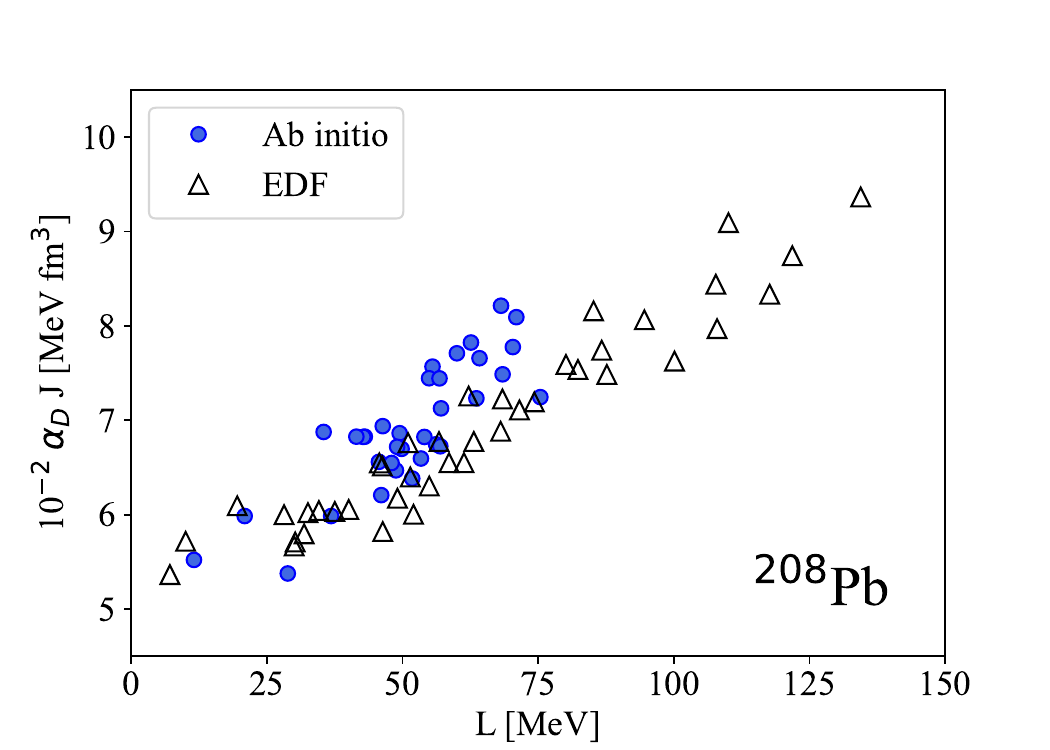}
\end{minipage}
 \caption{(Left) Correlation between the neutron-skin thickness of $^{208}$Pb and the slope parameter $L$ entering the nuclear-matter EOS from the ab initio calculations of Ref.~\cite{hu2022} and the relativistic and non-relativistic energy-density functional (EDF) models of Ref.~\cite{centelles2010}.
 (Right) Correlation between the product $\alpha_D J$ and the slope parameter $L$ from EDFs~\cite{rocamaza2013} and ab initio calculations~\cite{hu2022}.}
\label{fig:correlation-skin}
\end{figure}

A nucleus with a pronounced neutron skin is $^{208}$Pb, which has been investigated with different theoretical frameworks.
For a long time, both relativistic and non-relativistic energy-density functional (EDF) theory (see e.g. Ref.~\cite{schunck2019} and references therein) was the only practical approach to explore the heavy-mass region of the nuclear chart.
In recent years, however, advances in high-performance computing and many-body theory have enabled ab initio calculations to dramatically extend their reach in mass number~\cite{hergert2020}, culminating in the ab initio study of the neutron skin of $^{208}$Pb by Hu et al. in 2022~\cite{hu2022}. 
In both the EDF and ab initio approaches, a strong correlation between the neutron skin of $^{208}$Pb and the slope parameter $L$ emerges, as presented in Figure~\ref{fig:correlation-skin}. 

This correlation originates from the effect of symmetry pressure in neutron-rich matter: The parameter $L$ quantifies how rapidly the symmetry energy increases with density around saturation. 
As the pressure in symmetric matter vanishes at saturation density, $L$ is proportional to the pressure among neutrons. 
A larger $L$ then implies a higher neutron pressure around saturation, which pushes neutrons toward the nuclear surface and leads to a more pronounced neutron skin. 
Conversely, a smaller $L$ corresponds to a more compact distribution of neutrons and smaller neutron skin. 
This connection provides a valuable link between the properties of finite nuclei and infinite neutron matter, allowing neutron-skin measurements to constrain the density dependence of the symmetry energy and, by extension, the equation of state of neutron-rich matter relevant for neutron stars.

According to Eq.~(\ref{nskin}), measuring $R_{\rm skin}$ requires knowledge of both the proton and neutron distributions. 
Nuclear charge densities can be measured via electron scattering, enabling precise determinations of the proton distributions.
Neutron distributions, on the other hand, are much more difficult to constrain.
They have primarily been investigated using strongly-interacting probes (see e.g.,~\cite{hoffmann1980,zenihiro2010,aumann2017,klos2007}), but such measurements are affected by significant systematic uncertainties arising from the treatment of the strong force.  
Parity-violating elastic electron scattering (PVES), which probes the nucleus exclusively via the electroweak interaction, holds the promise of a reduced model dependence in the extraction of the neutron skin~\cite{thiel2019}, at the cost of less statistics.
PVES is applicable only to stable nuclei and  measures the parity-violating asymmetry $A_{PV}$, which is the difference between the cross sections of right-handed and left-handed electrons.
Under certain theoretical assumptions, this allows for the extraction of $R_{\rm skin}$~\cite{thiel2019, mammei2023}. 

In the last fifteen years, the Lead (Pb) Radius EXperiment (PREX) Collaboration at Jefferson Lab has provided the first experimental determination of the $^{208}$Pb neutron-skin thickness. In 2012, the first PREX experimental campaign (PREX-I~\cite{prex1}) found a neutron skin of $0.33^{+0.16}_{-0.18}$~fm. Later, in 2021, a second experiment was performed (PREX-II~\cite{prex}), where the accuracy was improved by approximately a factor of $2$, leading to a combined result of $R_{\rm skin} = 0.283\pm 0.071$~fm~\cite{prex}. We will label this result as PREX in the following. This result was translated, within EDFs, into a constraint on $L=106\pm 37$~MeV~\cite{reed2021}, a much higher value than favoured by previous constraints (see, e.g., Ref.~\cite{drischler2020}). Employing different families of EDFs, it was also found that the neutron skin of $^{208}$Pb shows a strong correlation with the skin of other nuclei, such as $^{48}$Ca~\cite{piekarewicz2012}. This has motivated a PVES measurement of the $^{48}$Ca neutron skin in the so-called Calcium Radius EXperiment (CREX), yielding $R_{\rm skin}=0.121\pm0.035$~fm~\cite{crex}, a much smaller value than for lead.

It is important to note that the connection between the measured $A_{PV}$ and $R_{\rm skin}$ is model dependent. 
Since $A_{PV}$ is measured at a single momentum transfer, $R_{\rm skin}$ is inferred through correlations between $A_{PV}$ and the weak radius or neutron skin obtained from nuclear models, which in the PREX and CREX analyses are EDFs~\cite{prex,crex}.
The implications of the PREX and CREX measurements are the subject of active debate in both the nuclear-physics and astrophysics communities. 
Refs.~\cite{reinhard2021,reinhard2022} used an EDF framework to directly consider the measured parity-violating asymmetry $A_{PV}$ in $^{48}$Ca and $^{208}$Pb instead of the corresponding $R_{\rm skin}$. 
They showed that the PREX and CREX $A_{PV}$ results cannot be simultaneously reproduced within one $\sigma$ with EDFs that ensure a good description of bulk properties of nuclei across the nuclear chart~\cite{reinhard2022}. 
Consequently, new functionals have been proposed to reconcile the differences between PREX and CREX~\cite{Reed:2023cap,Yue:2024srj,Kunjipurayil:2025xss}.
These models yield stiff equations of state that however fail to reproduce astrophysical constraints or lead to large nuclear-matter parameters that are inconsistent with other determinations.

It was claimed that certain quantum-electrodynamics (QED) corrections were missing from the analysis of the PREX/CREX data~\cite{Roca-Maza:2025vnr,Jakubassa-Amundsen:2025lgr}. 
However, it was shown that the proposed QED corrections were also incomplete, and when including all corrections, the results remain largely unchanged~\cite{Reed:2026bru}. 
More recently, No\"el et al.~\cite{noel2026} performed the first ab initio calculations of $A_{PV}$ for $^{48}$Ca and $^{208}$Pb and used the resulting correlations to infer $R_{\rm skin}$ from the measured asymmetries.
Using theoretically consistent inputs for charge and weak densities, they obtained $R_{\rm skin}=0.120(14)(12)$~fm for $^{48}$Ca, in very good agreement with CREX, and $R_{\rm skin}=0.187(25)(18)$~fm for $^{208}$Pb. 
The latter is considerably smaller but, given the large uncertainties, compatible with both PREX and the experiments based on strongly-interacting probes.

\begin{figure}[t]
\centering
    \includegraphics[scale=0.47]{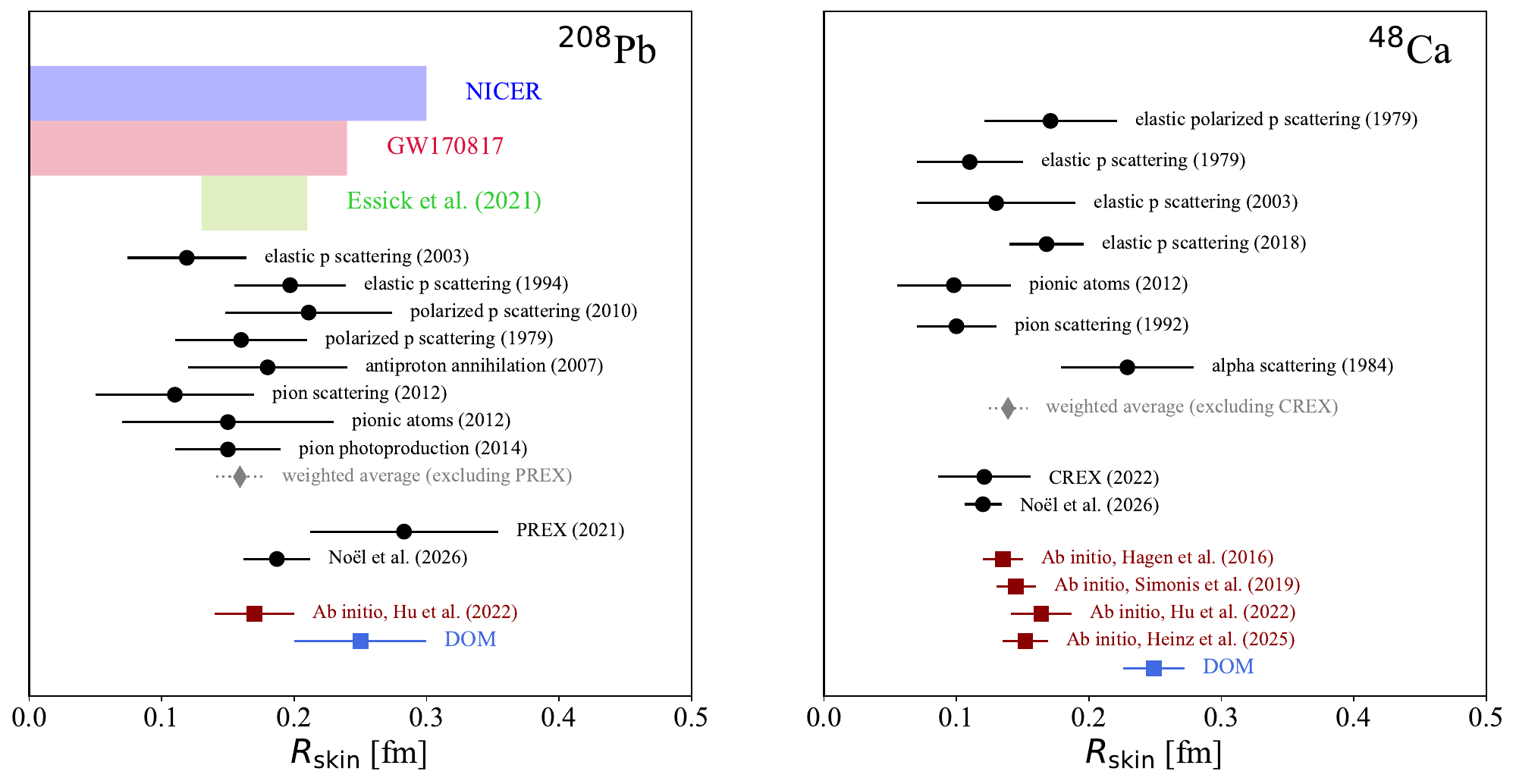}    
    \caption[Experimental determinations of the neutron-skin thickness of $^{48}$Ca and $^{208}$Pb, in comparison to ab initio and dispersive optical model calculations, together with constraints derived from gravitational waves and joint mass-radius neutron star measurements on $^{208}$Pb]
    {Experimental determinations (black) of the neutron-skin thickness of $^{208}$Pb (left panel) and $^{48}$Ca (right panel) (see Ref.~\cite{lattimer2023} and references therein), in comparison to ab initio~\cite{hagen2016,simonis2019,hu2022} (brown) and dispersive optical model~\cite{mahzoon2017,atkinson2020} (blue) calculations, and to the values inferred from the CREX and PREX-II asymmetries using ab initio correlations~\cite{noel2026}. 
    In the left panel, we also show comparisons with astrophysical constraints in the neutron-skin thickness from Refs.~\cite{fattoyev2018,reed2021,essick2021}.
    }
    \label{fig:exp-rskin}
\end{figure}

In Figure~\ref{fig:exp-rskin}, we show the available experimental data on the neutron skins of $^{208}$Pb and $^{48}$Ca, taken from Table 4 ($^{208}$Pb) and Table 5 ($^{48}$Ca) of Ref.~\cite{lattimer2023}, respectively (see Ref.~\cite{lattimer2023} and references therein for details on the experiments).
We compare those to ab initio results from Hagen et al.~\cite{hagen2016}, Simonis et al.~\cite{simonis2019} and Heinz et al.~\cite{heinz2025} for $^{48}$Ca, and from Hu et al.~\cite{hu2022} for both  $^{48}$Ca and $^{208}$Pb.
We also show the neutron skins inferred by No\"el et al.~\cite{noel2026} from the CREX and PREX-II asymmetries using ab initio correlations.
Results from the dispersive optical model (DOM) analysis of bound and scattering data~\cite{mahzoon2017,atkinson2020} are also reported for comparison. 
We also list numerical results in Table~\ref{tab:neutron_skin_comparison}.
In the case of $^{208}$Pb, we include EDF-based constraints obtained when using data from the gravitational-wave signal of GW170817~\cite{fattoyev2018}, as well as the joint mass-radius NICER measurements~\cite{reed2021}, and the value inferred by Essick et al.~\cite{essick2021}, on the basis of neutron-star observations and chiral EFT constraints. 

The CREX measurement for $^{48}$Ca is in excellent agreement with ab initio calculations from Refs.~\cite{hagen2016,simonis2019,hu2022,heinz2025}, while the DOM overestimates the experimental results. 
For $^{208}$Pb, the PREX result is more than $1\sigma$ larger than the average of previous $^{208}$Pb  measurements~\cite{lattimer2023}. 
While the DOM prediction agrees with PREX, ab initio calculations predict a smaller skin of $^{208}$Pb than PREX, in agreement with the average of previous measurements and with the value inferred from the PREX-II asymmetry using ab initio correlations~\cite{noel2026}. Hu et al.~\cite{hu2022} also observed that higher values of $R_{\rm skin}$ and $L$, in agreement with PREX, can be obtained in their calculations at the cost of degrading the description of nucleon-nucleon scattering data at low energies. From nuclear matter computations, Hu et al.~\cite{hu2022} obtained $38.3\leq L \leq 68.5$ MeV, supporting a softer EOS than inferred from PREX. Essick et al.~\cite{essick2021} find $L = 53^{+14}_{-15}$ MeV and infers values of $R_{\rm skin}$ in agreement with the ab initio results. The upper bounds obtained from NICER and GW170817 measurements, with the latter providing the more stringent limit, are consistent with all experimental determinations. 

The discussion above shows that more work on the theoretical assumptions entering the analysis of PVES experiments is needed. In this way, it will be possible to take full advantage of the forthcoming Mainz Radius Experiment (MREX)~\cite{mammei2023} at the MESA facility in Mainz, Germany, which is expected to substantially reduce the uncertainty in the PVES determination of the $^{208}$Pb neutron skin.

\begin{table}[t]
\centering
\caption{Comparison between theoretical predictions and  experimental determinations of the
neutron-skin thickness $R_{\mathrm{skin}}$ for $^{208}$Pb and $^{48}$Ca from the PREX and CREX experiments, respectively.
All values are given in fm.}
\label{tab:neutron_skin_comparison}
\begin{tabular}{llc}
\hline
Nucleus & Reference & $R_{\mathrm{skin}}$ [fm] \\
\hline
$^{208}$Pb
  & Ab initio~\cite{hu2022} & $0.170(30)$ \\
  & DOM~\cite{atkinson2020}                          & $0.250(50)$ \\
\cline{2-3}
  & PREX~\cite{prex}                  & $0.283(71)$ \\
  & PREX from ab initio correlations~\cite{noel2026}         & $0.187(25)(18)$ \\
\hline
$^{48}$Ca
  & Ab initio~\cite{hagen2016}   & $0.135(15)$ \\
  & Ab initio~\cite{simonis2019} & $0.145(15)$ \\
  & Ab initio~\cite{hu2022}     & $0.164(23)$ \\
  & Ab initio~\cite{heinz2025}   & $0.152(17)$ \\
  & DOM~\cite{mahzoon2017}                              & $0.249(23)$ \\
\cline{2-3}
  & CREX~\cite{crex}                     & $0.121(35)$ \\
  & CREX from ab initio correlations~\cite{noel2026}             & $0.120(14)(12)$ \\
\hline
\end{tabular}
\end{table}

\subsection{Electric dipole polarizability}
\label{alphaD-constraints}

An alternative experimental strategy to constrain the symmetry energy from finite nuclei is given by measuring the electric dipole polarizability.
In thermodynamics, the EOS of an ideal gas can be derived by investigating the behavior of the system when performing slow variations of volume, pressure, and temperature. 
Analogously, $J$ and $L$ can be studied in experiments where nuclei are slightly perturbed from their ground state by the presence of an external electromagnetic (EM) field. 
Owing to the small coupling constant, EM probes are especially convenient since perturbation theory can be applied and theoretical predictions can be directly compared with experiment.

When a nucleus is perturbed by an EM field, it is useful to introduce the response function (or strength function)
\begin{equation}
     R(\omega) = \sum_{\mu} \braket{\Psi_0|\Theta^{\dagger}|\Psi_{\mu}}\braket{\Psi_{\mu}|\Theta|\Psi_0}\delta(\omega - E_{\mu} + E_0)\,, 
     \label{response-eos-chapter}
\end{equation}
where $\Theta$ is the EM operator, $\ket{\Psi_0}$ and $\ket{\Psi_\mu}$ are the ground state and excited state $\mu$, respectively, and $E_{0}$ and $E_{\mu}$ are their energies.
The electric dipole, dominant at low energies, is the relevant EM operator for the purposes discussed here. 
It is defined as
\begin{equation}
    \Theta = e \sqrt{\frac{4\pi}{3}} \sum_{i=1}^A\; r_i\; Y_{10}(\hat{r}_i) \left(\frac{1+\tau^z_i}{2}\right)\,,
    \label{electric-dipole-operator}
\end{equation}
where $e$ is the electric charge, $r_i = |\vec{R}_i - \vec{R}_{\mathrm{CM}}|$ is the difference between the position of the $i$-th nucleon $\vec{R}_i$ and the center of mass $\vec{R}_{\mathrm{CM}}$, $Y_{10}(\hat{r}_i)$ is the spherical harmonic with rank $1$ and projection $0$, and $\tau^z_i$ is the $z$-axis projection of the $i$-th nucleon's isospin. 
Low photon energies excite the nucleus to discrete excited bound states, but when the energy overcomes the break-up threshold, i.e., the energy needed to remove a nucleon from the nucleus, the photon populates continuum excited states. 
In the continuum, the response function typically presents a giant dipole resonance, traditionally interpreted as a collective motion of all protons against all neutrons~\cite{eisenberg}, but if the nucleus has a large neutron excess, it may display a pygmy dipole resonances at low energies, due to the oscillations of the excess neutrons against a core.

Moments of the dipole response function provide insight into the dynamics of the nucleus and are often easier to calculate than the response function itself. 
In particular, the electric dipole polarizability $\alpha_D$, defined as an inverse energy-weighted sum rule of the dipole response function, is given by
\begin{equation}
    \alpha_D = 2\alpha \int_0^{\infty} d\omega \frac{R(\omega)}{\omega}\,,
    \label{alphaD-eos-chapter}
\end{equation}
and is useful to the purpose of linking finite nuclei to nuclear matter.
Experimentally, the dipole polarizability of stable nuclei can be extracted from measurements of photoabsorption cross sections, and many such experiments were conducted since the 70's-80's~\cite{ahrens1975,gurevich,ahrens1985, berman,ishkanov2002}. 
More recently, proton inelastic scattering $(p,p')$ at very forward angles~\cite{vnc2019} has been established as a solid experimental technique to access  $\alpha_D$ with good resolution, see, e.g., Refs.~\cite{tamii2011,hashimoto2015,birkhan2017,fearick2023}. 
Moreover, Coulomb excitation measurements in inverse kinematics made it possible to study $\alpha_D$ in neutron-rich nuclei such as $^{68}$Ni~\cite{rossi2013} and $^{6,8}$He~\cite{aumann1999,lehr2026}.

The connection between the electric dipole polarizability and the symmetry energy can be viewed through a simple analogy: the nucleus behaves like two overlapping fluids of protons and neutrons linked by a restoring “spring.” An external electric field displaces the two fluids and the symmetry energy determines the stiffness of the restoring force. 
A softer symmetry energy yields a smaller polarizability, while a stiffer one leads to a larger response. Thus, the dipole polarizability provides a sensitive probe of the symmetry energy and its density dependence.
Theoretical evidence of a correlation between $\alpha_D$ and the neutron-skin thickness $R_{\rm skin}$ has been first observed in the framework of EDF theory.  
Reinhard et al.~\cite{reinhard2010} found such correlation for $^{208}$Pb, and their results were then employed by Tamii et al.\cite{tamii2011} to infer a neutron skin estimate of $R_{\rm skin}=0.156^{+0.025}_{-0.021}$~fm from a new $(p,p')$ measurement of $\alpha_D=20.1\pm0.6$ fm$^3$.
This value is almost $50\%$ smaller than the PREX central value, but in agreement with the average of previous measurements based on strong probes~\cite{lattimer2023}, with ab initio calculations~\cite{hu2022} and with the re-analysis by Noël et al.~\cite{essick2021}. 
Subsequently, Piekarewicz et al.~\cite{piekarewicz2012} pointed out that while the correlation between $\alpha_D$ and $R_{\rm skin}$ emerges in an individual EDF family, the same linear dependence becomes less clean when combining a host of different models. 

In 2013, Roca-Maza et al.~\cite{rocamaza2013} observed a correlation between the product of $\alpha_D J$  and $R_{\rm skin}$ for $^{208}$Pb employing different sets of relativistic and non-relativistic EDFs. 
Such a correlation was explained via a droplet model~\cite{myers1974}, which represents a refinement of the liquid-drop model, from which the well-known von Weizsäcker mass formula describing the evolution of the binding energy per nucleon as a function of nucleon number $A$ is derived.
In that model, $\alpha_D$ depends on $A$ as~\cite{meyer1982}
\begin{equation}
    \alpha_D^{\mathrm{DM}} \propto  \frac{A\braket{R_m^2}}{48J} \left(1 + \frac{5}{3} \frac{9J}{4Q} A^{-1/3} \right)\,,
    \label{alphaD-droplet}
\end{equation}
where $R_m = \braket{R_m^2}^{1/2}$ is the matter radius of the nucleus and $Q$ is the surface stiffness coefficient, quantifying the resistance of neutrons against being separated by protons~\cite{myers1974}. 
Ref.~\cite{rocamaza2013} showed that Eq.~(\ref{alphaD-droplet}) can be expressed in terms of the neutron-skin thickness $R_{\rm skin}^{\mathrm{DM}}$ as
\begin{equation}
    \alpha_D^{\mathrm{DM}} \propto \frac{A\braket{R_m^2}}{48J} \left(1 + \frac{5}{2} \frac{R_{\rm skin}^{\mathrm{DM}}A}{ R_m (N-Z)}\right)\,,
\end{equation}
which indicates that the product $\alpha_D J$, instead of $\alpha_D$ alone, correlates with $R_{\rm skin}$ and therefore $L$. 
The $\alpha_D J$–$L$ correlation for $^{208}$Pb obtained within EDF in Ref.~\cite{rocamaza2013} is shown in Fig.~\ref{fig:correlation-skin}, along with ab initio results from Hu et al.~\cite{hu2022}, which later confirmed this correlation within a different framework.
It is important for assessing the robustness of such a correlation that it naturally arises in very different theoretical frameworks. 
Therefore, an experimental value of $\alpha_D$ can be exploited to get a direct relation between $J$ and $L$, if used in conjunction with theoretical calculations.

By combining the experimental determination of $\alpha_D$ in $^{208}$Pb~\cite{tamii2011} with a realistic value of $J= 32\pm2$ MeV~\cite{lattimer2013}, Roca-Maza et al.~\cite{rocamaza2013} obtained $R_{\rm skin}(^{208}$Pb$) = 0.165\pm 0.043$ fm$^3$ and $L=43\pm 26$ MeV. This value is consistent with Noël et al.~\cite{noel2026} as well as with the ab initio calculations of Hu et al.~\cite{hu2022}, but lies below the PREX result. Subsequently in Ref.~\cite{rocamaza2015}, Roca-Maza et al.~expanded their previous analysis by including also new experimental data for the polarizabilities of $^{68}$Ni~\cite{rossi2013} and $^{120}$Sn~\cite{hashimoto2015}.
They found a combined result of $30\leq J\leq 35 \;\mathrm{MeV}$ and $20\leq L\leq 66 \;\mathrm{MeV}$. These ranges are consistent with Essick et al.~\cite{essick2021} and of Hu at al.~\cite{hu2022}. The constraints coming from the polarizability then indicate a softer EOS with respect to PREX.

In recent years, new developments in ab initio nuclear theory made it possible to calculate  $\alpha_D$ for light and medium-mass nuclei.
Computing $\alpha_D$ is technically more arduous than computing $R_{\rm skin}$, because, as shown in Eq.~(\ref{alphaD-eos-chapter}), $\alpha_D$ depends on the response function, and consequently, on the excited states of the nucleus.
Combining the Lorentz Integral Transform (LIT)~\cite{efros2007}, a few-body technique allowing for a correct treatment of continuum states, with the mild computational scaling of coupled-cluster (CC) theory~\cite{hagen2014}, the so-called LIT-CC method~\cite{bacca2013,bacca2014} was developed. This method was successfully tested on $^4$He, $^{16}$O, and $^{22}$O and found results in good agreement with available experimental data~\cite{miorelli2016}. 
LIT-CC calculations of the dipole polarizability of $^{48}$Ca were provided by employing different chiral EFT models, leading to $\alpha_D$ values in the range of $[2.19-2.60]$~fm$^3$~\cite{hagen2016} obtained by exploiting the correlations of $\alpha_D$ with various EM observables. 
This range is consistent with the experimental value of $\alpha_D=2.07(22)$~fm$^3$ determined via a $(p,p')$ measurement~\cite{birkhan2017}, and with EDF results. Later on, by increasing the accuracy of the LIT-CC method, more precise estimates of the $^{48}$Ca polarizability were provided in Refs.~\cite{miorelli2018,simonis2019}, leading to an $\alpha_D$-range of $[1.92-2.38]$ fm$^3$. In Table~\ref{tab:alphaD_48Ca}, we present all the above mentioned results for $\alpha_D$ and compare them to the experimental value of Ref.~\cite{birkhan2017}.
\begin{table}[t]
    \centering
    \caption{Comparison between theoretical predictions of the electric dipole polarizability $\alpha_D$ of
    $^{48}$Ca and experiment.\\}
    \label{tab:alphaD_48Ca}
    \begin{tabular}{ll}
        \hline
        Method & $\alpha_D$ [fm$^3$] \\
        \hline
        Coupled-cluster~\cite{hagen2016}  & $2.40 \pm 0.20$ \\
        Coupled-cluster~\cite{simonis2019} & $2.15 \pm 0.23$ \\
        IMSRG~\cite{hu2022}      & $2.30^{+0.31}_{-0.26}$ \\\\
        EDF~\cite{rocamaza2015}             & $2.306 \pm 0.089$ \\
        EDF~\cite{piekarewicz2012}           & $2.30 \pm 0.10$ \\
        Experiment~\cite{birkhan2017} &     $1.85$--$2.29$ \\
        \hline
    \end{tabular}
\end{table}

In general, we observe an excellent agreement between ab initio predictions, EDF theory and experiment, but one finds that the improved LIT-CC results of Ref.~\cite{simonis2019} lead to a smaller $\alpha_D$ with respect to EDF calculations, in better agreement with experiment. 

More recently, the same set of chiral interactions from Refs.~\cite{hagen2016,simonis2019}, together with the LIT-CC method, was employed to predict the dipole polarizability of $^{40}$Ca, measured at RCNP Osaka via inelastic ($p$,$p'$) scattering~\cite{fearick2023}. The range of theoretical LIT-CC predictions, $1.35 \lesssim \alpha_D \lesssim 2.05$ fm$^3$, is consistent with the experimental value of $1.92(17)$ fm$^3$ and with corresponding EDF results. In particular, it was found that the chiral interaction NNLO$_{\rm sat}$, as well as EDFs fitted on the dipole polarizability of $^{208}$Pb, were able to reproduce both the dipole polarizability of $^{40}$Ca and $^{48}$Ca.

The LIT-CC method was also utilized to tackle the unstable neutron-rich nucleus $^{68}$Ni in Ref.~\cite{kaufmann2020}. An $\alpha_D$ value in the range $[3.58-4.65]$ fm$^3$ was obtained, in good agreement with the experimental value measured by Rossi et al.~\cite{rossi2013} and refined by Roca-Maza et al.~\cite{rocamaza2015} of $\alpha_D=3.40(23)$ fm$^3$ . 
The polarizability was also calculated with the ab-initio self-consistent Green’s function method by Raimondi and Barbieri~\cite{raimondi2019}, where $\alpha_D = 3.60$ fm$^3$ was obtained using the NNLO$_{\rm sat}$ interaction.
With the same interaction, the LIT-CC yields a larger value of $\alpha_D = 4.65(49)$ fm$^3$. 
Most likely, the discrepancy is due to the different many-body methods and convergence parameters. The LIT-CC calculations presented in Ref.~\cite{kaufmann2020} yield a neutron-skin thickness of $^{68}$Ni in the range $[0.18 - 0.20]$~fm,  in agreement with the EDF analysis of Roca-Maza et al.~\cite{rocamaza2015}. 

The dipole polarizability of nuclei as heavy as $^{208}$Pb has been recently computed with ab initio methods by Hu et al.~\cite{hu2022}, as already shown in Figure~\ref{fig:correlation-skin}.
In particular, the in-medium similarity renormalization group IMSRG~\cite{hergert2016} method was employed: the non-implausible chiral interactions predicted $\alpha_D = 22.6^{+2.6}_{-1.8}$ fm$^3$, in agreement with the experimental value by Tamii et al.~\cite{tamii2011}.

Ab initio calculations are now capable of reproducing both the experimental neutron-skin thickness from CREX~\cite{crex} and the dipole polarizability of $^{40,48}$Ca~\cite{birkhan2017,fearick2023} as well as that of $^{68}$Ni~\cite{kaufmann2020} and $^{208}$Pb~\cite{hu2022}. 
The nuclear force models used in Refs.~\cite{hagen2016,simonis2019,fearick2023,kaufmann2020}, when implemented in infinite nuclear-matter calculations, yield~\cite{hagen2016,ekstrom2015}
$27\leq J\leq 33\;\mathrm{MeV}$ and 
    $44\leq L\leq 49\;\mathrm{MeV}\,$. Ref.~\cite{hu2022}, based on a set of non-implausible chiral interactions, finds $J = 31.1\pm ^{2.1}_{2}$ and $L = 52.7^{+15.8}_{-14.4}$ MeV. 

\begin{figure*}[t]
\centering
    \includegraphics[scale=0.49]{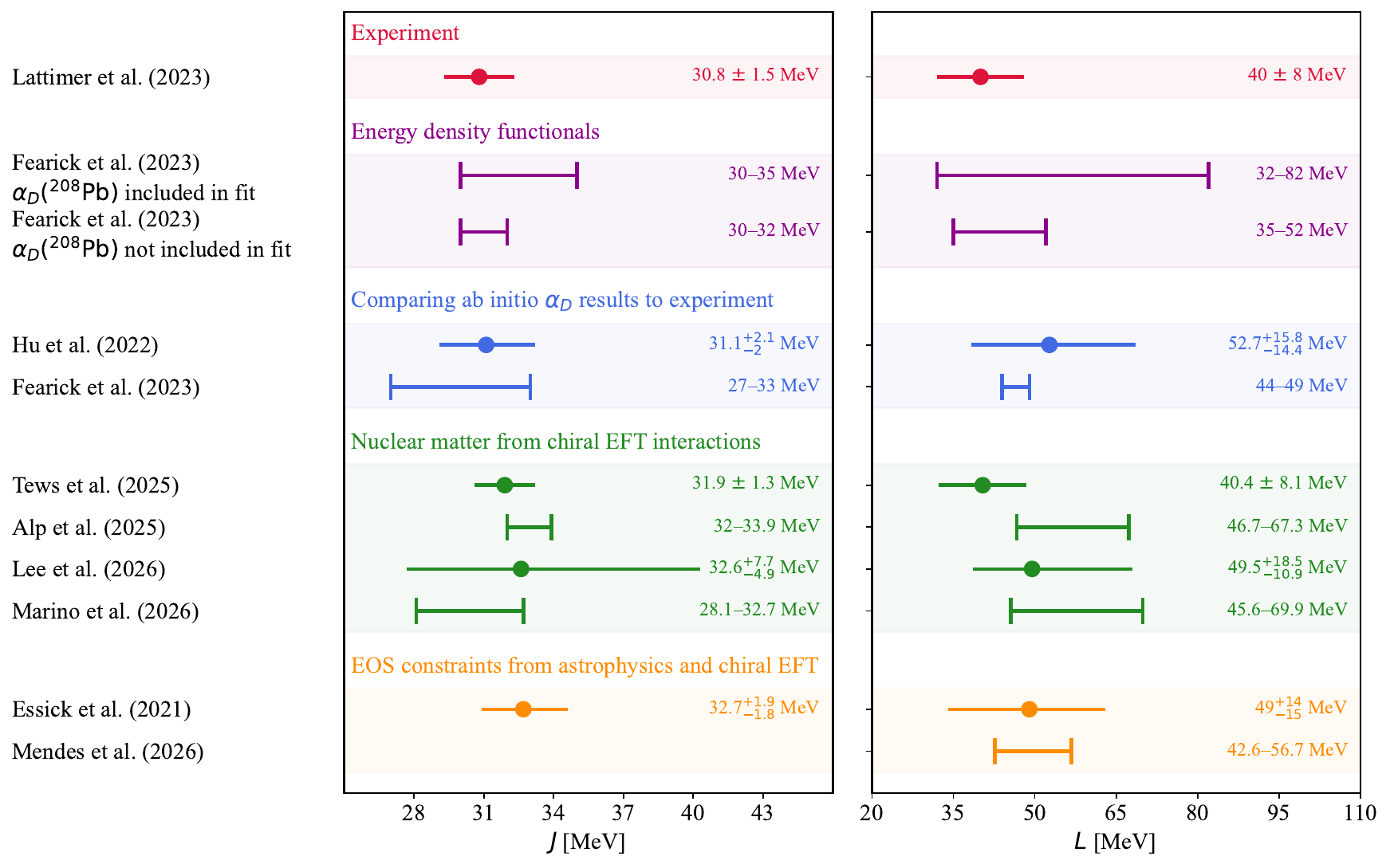}    
    \caption{Constraints on the symmetry energy $J$ (left panel) and slope parameter $L$ (right panel) obtained starting from the chiral interactions employed in Ref.~\cite{hu2022,fearick2023},  compared with the global analysis of Lattimer~\cite{lattimer2023}, the EDF ranges of Ref.~\cite{fearick2023}, the nuclear matter calculations of Tews et al.~\cite{tews2025}, Alp et al.~\cite{alp2025}, Lee et al.~\cite{lee2026} and Marino et al.~\cite{marino2026} and the values inferred by Essick et al.~\cite{essick2021, essick2021_prc} and Mendes et al.~\cite{mendes2026}. 
    For Ref.~\cite{alp2025,marino2026}, the range shown includes results obtained with different sets of chiral interactions. For Ref.~\cite{mendes2026}, the constraint on L encompasses the range obtained with piecewise-polytrope and speed-of-sound high-density extension. See text for details.}
    \label{fig:lconstraints}
\end{figure*}

Let us compare such constraints to recent results reported in the literature in Figure~\ref{fig:lconstraints}. We focus first on the review by Lattimer~\cite{lattimer2023}, where a weighted average of experimental data for the neutron skin of $^{48}$Ca and $^{208}$Pb was used in conjunction with theory to obtain allowed ranges for symmetry energy parameters. We also consider the two EDF ranges from Ref.~\cite{fearick2023}, differing in the inclusion of the polarizability of $^{208}$Pb in the EDF fit. In addition, Figure~\ref{fig:lconstraints} reports a representative set of results from nuclear matter calculations~\cite{tews2025,alp2025,lee2026,marino2026}. The latter are based on both local~\cite{tews2025} and non-local~\cite{alp2025,lee2026,marino2026} interactions, and on different many-body methods (Quantum Monte Carlo~\cite{tews2025}, Many-Body Perturbation Theory~\cite{alp2025,lee2026}, Self-Consistent Green's Function~\cite{marino2026}). Constraints informed by both astrophysical observations and chiral EFT computations~\cite{essick2021} and~\cite{mendes2026} are also included for comparison.

The constraints obtained using the chiral interactions of Refs.~\cite{hu2022,fearick2023} and the EDFs of Ref.~\cite{fearick2023} are consistent with the nuclear matter calculations of Refs.~\cite{tews2025,alp2025,lee2026,marino2026}. They also agree with the ranges inferred by combining chiral EFT with neutron star observations~\cite{essick2021,mendes2026} and with the global analysis of Lattimer~\cite{lattimer2023}. Across these results, $J$ generally lies around $30-32$ MeV, while allowed ranges of $L$ are broader, typically spanning $35-70$ MeV. Constraints on $L$ remain below the range inferred from PREX~\cite{reed2021}, even when correlated uncertainties in nuclear matter calculations are taken into account~\cite{lee2026}. As for the EDF results of Ref.~\cite{fearick2023},  restricting the comparison to functionals fitted to the polarizability of $^{208}$Pb substantially narrows the predicted range of $L$, bringing it into closer agreement with the chiral EFT constraints.

\section{CONSTRAINTS ON THE INCOMPRESSIBILITY}
\label{k-constraints-section}
Electromagnetic observables can be used to constrain other parameters of the nuclear-matter EOS. 
In this section, we focus on the incompressibility $K_0$, which determines the curvature of $E/N$ in symmetric nuclear matter at saturation density, see Eq.~(\ref{incompressibilitysnm}).
This parameter can be constrained from the isoscalar monopole response in finite nuclei~\cite{blaizot1980}. This corresponds to Eq.(\ref{response-eos-chapter}) where the EM operator is the isoscalar monopole,
\begin{equation}
    \Theta = \frac{1}{2} \sum_{i=1}^A r_i^2\,, 
    \label{monopole-operator}
\end{equation}
written with the same notation used for Eq.~(\ref{electric-dipole-operator}). 

Early measurements of isoscalar monopole responses in the 1970s~\cite{youngblood1977} have shown that the monopole strength is dominated by a broad peak, known as isoscalar giant monopole resonance (ISGMR),  traditionally interpreted as a ``breathing" mode. 
In 1980, Blaizot~\cite{blaizot1980} established the following relation between the average energy of the ISGMR, $E_{\mathrm{ISGMR}}$, and the incompressibility of an A-body finite nucleus
\begin{equation}
    K_A = \frac{M}{\hbar^2} \braket{R_m^2} E^2_{\mathrm{ISGMR}}\,, 
\end{equation}
where $M$ is the mass of the nucleon.
The average energy $E_{\mathrm{ISGMR}}$ can be estimated as
\begin{equation}
    \widetilde{E}_{\mathrm{ISGMR}} = \frac{m_1}{m_0}\,,
    \bar{E}_{\mathrm{ISGMR}} = \sqrt{\frac{m_k}{m_{k-2}}}\,,
\label{isgmrk}
\end{equation}
where $m_k$ are  moments of the monopole response function, defined as
\begin{equation}
    m_k = \int d\omega \omega^{k} R(\omega)\,.
\end{equation}
The spread between the different expressions in Eq. (\ref{isgmrk}) quantifies the fragmentation of the monopole strength distribution. 

Once $K_A$ values are available for different nuclei, one can consider the following expansion for $K_A$~\cite{blaizot1980,garg2018}
\begin{equation}
    K_A = K_{\rm vol} + K_{\rm surf} A^{-1/3} + K_{\rm Coul} Z^2 A^{-4/3} + K_{\tau} \frac{(N-Z)^2}{A^2},
    \label{leptodermousexp}
\end{equation}
analogous to the case of the liquid drop formula for the binding energy per nucleon. 
Here $K_{\rm vol}$, $K_{\rm surf}$, $K_{\rm Coul}$, $K_{\tau}$ are the volume, surface, Coulomb and isospin coefficients, respectively.
Eq.~(\ref{leptodermousexp}) is often referred to as a leptodermous expansion for $K_A$. 
In the $A\rightarrow\infty$ limit, the volume term $K_{\rm vol}$ can be interpreted as the incompressibility of infinite nuclear matter $K_0$. 
Therefore, assuming $K_{\rm vol}|_{A\rightarrow \infty}=K_0$, when having a sufficient number of experimental data or theoretical calculations of $E_{\mathrm{ISGMR}}$ at disposal,  Eq.~(\ref{leptodermousexp}) can be fitted, and a $K_0$ value can be extracted starting from properties of finite nuclei. 

Presently, the extracted $K_0$ values carry large uncertainties~\cite{garg2018,rocamaza2018}. Eq.~(\ref{leptodermousexp}) has been employed to constrain $K_0$ starting from the available experimental data in Ref.~\cite{stone2014}, leading to a broad range: $250 < K_0 < 315$ MeV. 
Moreover, $K_0$ can be inferred experimentally from heavy-ion collision data, such as those obtained by the FOPI Collaboration at GSI~\cite{lefevre2016,wang2018}, by comparing the experimentally extracted collective flow at different beam energies to transport model simulations.
Heavy-ion collision data favor softer values for $K_0$, however uncertainties remain large. 
Depending on the transport model used, the same Au + Au collision data between $0.4$ and $1.5$ GeV/nucleon from FOPI led to a $K_0$ constraint of $190 \pm 30$ MeV and $220 \pm 40$ MeV in Refs.~\cite{lefevre2016} and~\cite{wang2018}, respectively. 
A more recent Bayesian analysis~\cite{wei2026} led to $K_0 = 234.9^{+59.6}_{-63.6} \;(226.7^{+55.7}_{-50.4})$ MeV for a beam energy of $0.25$ ($0.4$)~GeV/nucleon. 

On the theoretical side, the connection between the incompressibility of nuclear matter and isoscalar monopole responses has been first employed in a indirect way in EDF calculations. 
Shlomo et al.~\cite{shlomo2006} compared the available experimental data on the ISGMR of $^{208}$Pb and $^{90}$Zr to EDF calculations. 
By selecting the EDFs capable of reproducing experimental data for $E_{\mathrm{ISGMR}}$, the value of $240\pm20$~MeV for the incompressibility of nuclear matter was identified.

\begin{figure}[t]
  \centering
  \includegraphics[width=0.8\textwidth]{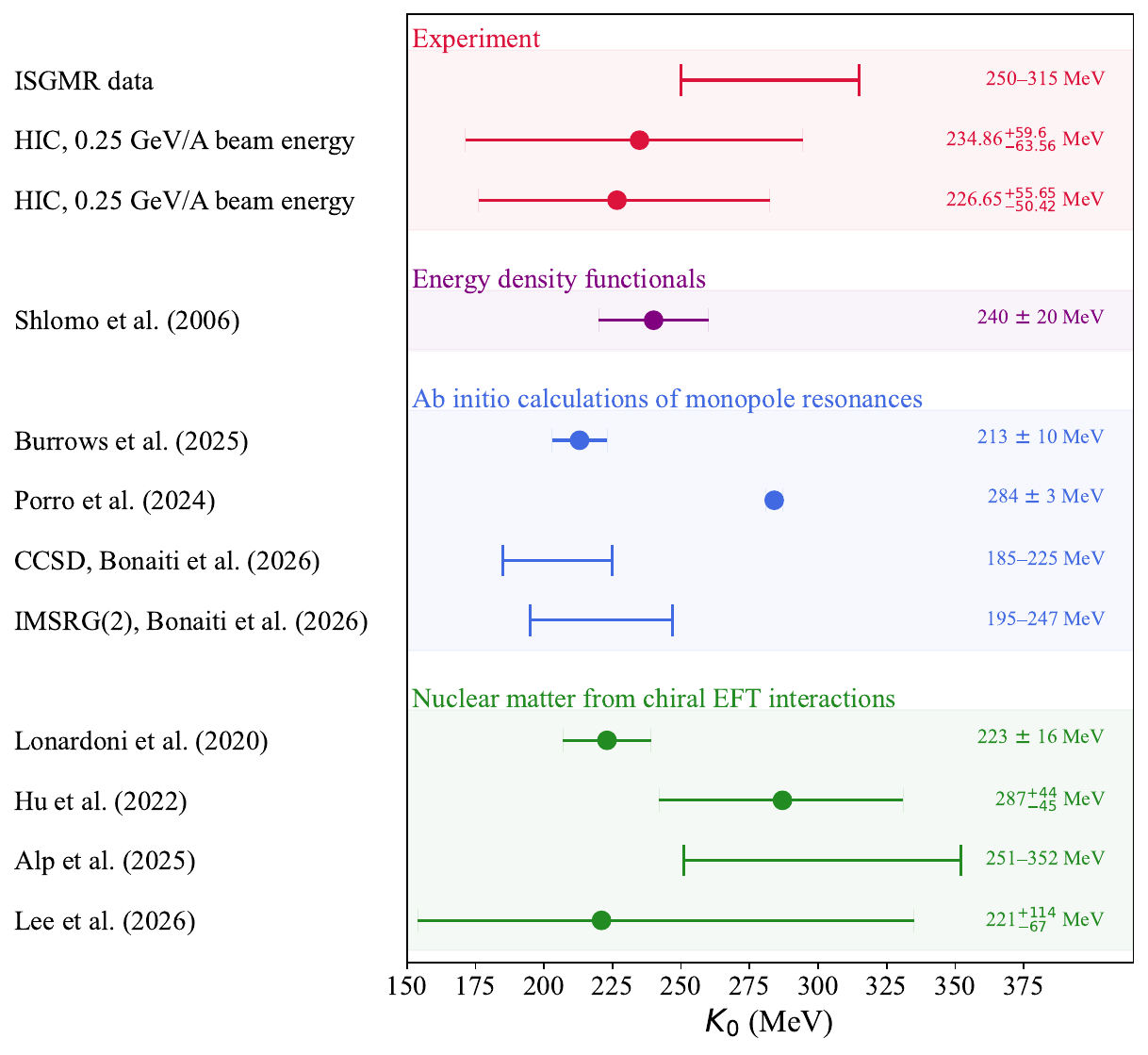}
  \caption{Comparison between nuclear incompressibility constraints extracted from experiment, ab initio calculations of finite nuclei, nuclear matter computations based on chiral EFT interactions and energy density functionals. Experimental extractions have been carried out based on isoscalar monopole resonance data~\cite{stone2014} and heavy-ion collisions~\cite{wei2026}. Extractions based on ab initio calculations and Eq.(\ref{leptodermousexp}) are taken from Refs.~\cite{burrows2023,porro2024,bonaiti2025b}. Ranges from nuclear matter computations are obtained from Ref.~\cite{lonardoni2020,hu2022,alp2025,lee2026}. The EDF range is taken from Ref.~\cite{shlomo2006}.}
  \label{fig:incompressibility}
\end{figure}

Since the liquid-drop model provides only a phenomenological description of nuclei, Eq.~(\ref{leptodermousexp}) should be interpreted with caution. 
However, it may be interesting to compare extractions of $K_0$ starting from calculations of finite nuclei and using Eq.~(\ref{leptodermousexp}) against results obtained from nuclear matter calculations~\cite{garg2018}. 
Recently, several ab initio studies have attempted to connect finite-nucleus monopole responses to $K_0$. Ref.~\cite{burrows2023} analyzed the $^4$He, $^{16}$O, $^{20}$Ne and $^{40}$Ca nuclei with the NNLO$_{\rm opt}$ and NNLO$_{\rm sat}$ interactions using the symmetry-adapted no-core shell model framework.
Ref.~\cite{porro2024} used the projected generator coordinate method to investigate the $^{24}$Mg, $^{28}$Si and $^{46}$Si nuclei employing the chiral NN+3N interaction from~\cite{huther2020}.  Finally, Ref.~\cite{bonaiti2025b} studied the $^{16}$O, $^{40}$Ca, $^{56}$Ni and $^{100}$Sn nuclei using  CC and IMSRG and the NNLO$_{\rm sat}$ and $\Delta$NNLO$_{\rm GO}$(394) interactions. 
 In all these works, the average energy $\sqrt{m_1/m_{-1}}$ has been used to estimate $E_{\rm ISGMR}$. 
Moreover, extracted values for $K_0$ have been obtained by a fit to a reduced version of Eq.~(\ref{leptodermousexp}), including only the $K_{\rm vol}$ and $K_{\rm surf}$ terms. 
Results for the incompressibility of nuclear matter extracted from these ab initio calculations are shown in Figure~\ref{fig:incompressibility}, along with the other mentioned results. A representative set of nuclear matter results, discussed already in the previous Section, is also reported. The Quantum Monte Carlo estimate in this case comes from Lonardoni et al.~\cite{lonardoni2020}.

The incompressibility extracted by Bonaiti et al.~\cite{bonaiti2025b} is consistent with constraints from heavy-ion collision data, phenomenological EDF ranges, and the extraction by Burrows et al.~\cite{burrows2023}. However, it lies below both the extrapolation from ISGMR data reported in Ref.~\cite{stone2014} and the result of Porro et al.~\cite{porro2024}. Compared with nuclear matter calculations, the constraint of Bonaiti et al.~\cite{bonaiti2025b} agrees with the predictions of Refs.~\cite{lonardoni2020,lee2026} and lies at the lower end of the ranges reported in Refs.~\cite{hu2022,alp2025}. The constraint obtained with local interactions~\cite{lonardoni2020} tends to favor lower values than those obtained with non-local interactions~\cite{hu2022,alp2025}. The range reported by Lee et al.~\cite{lee2026}, which accounts for correlated uncertainties, overlaps with all nuclear matter constraints.

Differences also arise between finite-nuclei and infinite-matter constraints when the underlying interactions are held fixed. Bonaiti et al.~\cite{bonaiti2025b} found that the values of $K_0$ extrapolated from ab initio calculations of ISGMR average energies were lower than those obtained from nuclear matter calculations using the same chiral interactions. The latter results are also included in the ranges reported in Refs.~\cite{hu2022,alp2025}. Two methodological differences are relevant to this comparison. First, nuclear matter calculations omit the Coulomb interaction, which are included in ab initio calculations of monopole resonances. Second, nuclear matter calculations determine $K_0$ from the curvature of the energy per particle, $E/A$, with respect to density at the predicted saturation point. A saturation point outside the empirical range can therefore affect the resulting value of $K_0$. The extraction from ab initio ISGMR calculations also has limitations. It relies on a small set of relatively light nuclei ($A\leq 100$) with $N=Z$, restricting the fit to Eq.~(\ref{leptodermousexp}) to the determination of $K_{\rm vol}$ and $K_{\rm surf}$. Extending the calculations to a broader set of nuclei would help assess the robustness of the current estimate.

\section{CONCLUSIONS} \label{conclusions}
The study of finite nuclei provides a powerful and complementary avenue for constraining the nuclear matter equation of state around saturation density. 
As discussed in this review, observables such as neutron-skin thicknesses of atomic nuclei, their electric dipole polarizabilities, and isoscalar monopole responses probe different aspects of the EOS and, when confronted with ab initio calculations, provide constraints on the symmetry energy and the incompressibility of nuclear matter.

Further progress will require improvements on both the experimental and theoretical sides. 
Parity-violating electron scattering is considered to be a clean probe of neutron distributions, but the extraction of neutron-skin thicknesses from the measured parity-violating asymmetries is not entirely free from theoretical assumptions and model dependence.
Recent developments have highlighted the importance of carefully quantifying these effects. 

At the same time, reducing the experimental uncertainties remains essential, and forthcoming measurements such as MREX in Mainz are expected to provide significantly more precise information on the neutron distribution, and hence neutron-skin thickness, of $^{208}$Pb. 
More generally, improved experimental precision for all the observables discussed here will directly enhance their ability to discriminate among different predictions for the EOS.
Additional measurements of electric dipole polarizabilities will also be particularly valuable. 
Extending the available data to isotopic chains and to increasingly neutron-rich systems will provide new tests of the correlations between finite-nucleus observables and the symmetry energy and will allow increasingly stringent comparisons with ab initio calculations. 
Similar considerations apply to monopole observables, for which present extractions of the nuclear incompressibility still carry sizable uncertainties.

On the theory side, substantial effort has been devoted in recent years to uncertainty quantification, including uncertainties associated with nuclear interactions, many-body approximations, and the modeling of the reactions and processes used to extract nuclear observables. 
Continuing this effort will be crucial if increasingly precise experimental data are to be fully exploited.

Finally, future astrophysical observations of neutron stars and neutron-star mergers will continue to provide increasingly stringent and complementary information on the EOS. The combination of terrestrial experiments, ab initio nuclear theory, and astrophysical observations thus offers the most promising path toward a quantitatively controlled understanding of nuclear matter.

\section*{DISCLOSURE STATEMENT}
 The authors are not aware of any affiliations, memberships, funding, or financial holdings that
might be perceived as affecting the objectivity of this review. 

\section*{ACKNOWLEDGMENTS}
This work was supported by the Deutsche
Forschungsgemeinschaft (DFG)
through the 
Cluster of Excellence ``Precision Physics, Fundamental
Interactions, and Structure of Matter" (PRISMA$^+$ EXC 2118/1, Project ID 390831469) and through the Collaborative Research Center ``Hadron and Nuclei as discovery tools" (Project ID 514321794), and through Project-ID
279384907 – SFB 1245 and by the U.S. Department of Energy, Office of Science, Office of Nuclear Physics, under the FRIB Theory Alliance Award No. DE-SC0013617, by the Office of Science, Office of Nuclear Physics program under Award Number DE-SCL000001, and by the Office of Advanced Scientific Computing Research and Office of Nuclear Physics, Scientific Discovery through Advanced Computing (SciDAC) program
(SciDAC-5 NUCLEI). 
This work was also supported by the U.S. Department of Energy through Los Alamos National Laboratory (LANL). 
LANL is operated by Triad National Security, LLC, for the National Nuclear Security Administration of U.S. Department of Energy (Contract No.~89233218CNA000001).
This work benefited from discussions at the Institute for Nuclear Theory at the University of Washington during the program INT 26-1: ``Nuclear Hamiltonians for Advancing Nuclear Physics and Beyond'', which was supported in part by the INT's U.S. Department of Energy grant No. DE-FG02-00ER41132.


\section*{LIST OF ACRONYMS}
\begin{longtable}{p{4cm}p{10cm}}
\textbf{3N(F)}					    & Three-nucleon (force)\\
\textbf{CC}							& Coupled-cluster\\
\textbf{CREX}						& Calcium radius experiment\\
\textbf{DOM}                    &Dispersive optical model\\
\textbf{EFT}					& Effective field theory\\
\textbf{EDF}							& Energy density functional\\
\textbf{EM}                         & Electromagnetic or in the context of forces, Entem and Machleidt \\
\textbf{EOS}							& Equation of state\\
\textbf{IMSRG}         & In-medium similarity renormalization group\\
\textbf{ISGMR}         & Isoscalar giant monopole resonance \\
\textbf{LEC}							& Low-energy constant\\
\textbf{LIT}							& Lorentz integral transform\\
\textbf{LIT-CC}						& Lorentz integral transform coupled-cluster\\
\textbf{LO}							& Leading order\\
\textbf{MESA}                       & Mainz energy-recovering superconducting accelerator \\
\textbf{MREX}						& Mainz radius experiment\\
\textbf{NCSM}                       & No-core shell model \\
\textbf{N$^3$LO}						& Next-to-next-to-next-to-leading order\\
\textbf{N$^2$LO} or \textbf{NNLO}   & Next-to-next-to-leading order\\
\textbf{NICER}                      & Neutron star interior composition explorer \\
\textbf{NN}							& Nucleon-nucleon \\
\textbf{PGCM}                       & Projected generator coordinate method\\
\textbf{PNM}                        & Pure neutron matter\\
\textbf{PREX}                       & Lead ($^{208}$Pb) radius experiment\\
\textbf{PVES}                       & Parity violating electron scattering\\
\textbf{RPA}                        & Random phase approximation\\
\textbf{QCD}						& Quantum chromodynamics\\
\textbf{SCGF}                       & Self consistent Green's function\\
\textbf{SA-NCSM}                       & Symmetry-adapted no-core shell model \\
\textbf{SNM}                        & Symmetric nuclear matter\\
\textbf{SRG}                        & Similarity renormalization group \\
\textbf{TOV}                        & Tolman-Oppenheimer-Volkoff (equations)
\end{longtable}
\section*{LITERATURE\ CITED}
\bibliographystyle{ar-style5}
\bibliography{bibliography_all}

\end{document}